\documentclass[AMA,STIX1COL]{WileyNJD-v2}
\usepackage{verbatim}

\articletype{Research article}%

\def\m#1{\mathsf{#1}} 

\newcommand{\cl}[2]{\ensuremath{\mathit{Cl}_{#1,#2}}}

\DeclareMathOperator{\Det}{Det} 
\DeclareMathOperator{\Tr}{Tr}

\newcommand{\bbR}{\ensuremath{\mathbb{R}}}

\newcommand{\ii}{\mathrm{i}}

\newcommand{\ee}{\mathrm{e}} 
\newcommand{\dd}{\mathrm{d}} 

\def\m#1{\mathsf{#1}}

\def\e#1{\mathbf{e}_{#1}} 

\usepackage{color}
\newcommand{\mycomment}[1]{} 

\begin{document}

\title{Multivector (MV) functions in Clifford algebras of arbitrary dimension: Defective MV case}

\author[1]{A. Acus*}
\author[2]{A. Dargys}


\address[1]{\orgdiv{Institute of Theoretical Physics and Astronomy}, \orgname{Vilnius University}, \orgaddress{\state{Saul{\.e}tekio 3, LT-10257 Vilnius}, \country{Lithuania}}}

\address[2]{\orgdiv{Semiconductor Physics Institute}, \orgname{Center for Physical Sciences and Technology}, \orgaddress{\state{Saul{\.e}tekio 3, LT-10257 Vilnius}, \country{Lithuania}}}

\corres{*A. Acus, Institute of Theoretical Physics and Astronomy, Vilnius University, Saul{\.e}tekio 3, LT-10257 Vilnius. \email{arturas.acus@tfai.vu.lt}}


\abstract[Summary]{Explicit formulas to calculate MV functions in
a basis-free representation are presented for an arbitrary Clifford geometric algebra \cl{p}{q}.
The formulas  are based on analysis of the roots of minimal MV
polynomial and covers defective MVs, i.e. the MVs that have non-diagonalizable
matrix representations. The method may be generalized
straightforwardly to matrix functions and to finite dimensional
linear operators. The results can find wide application in
Clifford algebra analysis.}


\jnlcitation{\cname{%
\author{A. Acus},
 and
\author{A. Dargys}} (\cyear{2024}),
\ctitle{ Multivector (MV) functions in Clifford algebras of arbitrary dimension: Defective MV case}, \cjournal{Math Meth Appl Sci.},
\cvol{2022;XX:x--xx.}.}

\maketitle

\footnotetext{\textbf{Abbreviations:} MV, multivector; GA, geometric (Clifford) algebra; nD, n dimensional vector space}

\section{Introduction}

Computation of matrix functions often arises as an important step in problems related to
physical, economical, biological, etc processes. In geometric algebra (GA), this is equivalent to computation of function of a
multivector (MV)~\cite{Gurlebeck1997,Lounesto1997,Marchuk2020}. A vast topic on matrix functions is well covered in the 
book~\cite{Higham2008}.

If matrix is non-defective (diagonalizable), a typical computation step makes a start with a diagonalization procedure. Then, the
function of diagonal matrix is straightforward to compute. In the non-diagonalizable case the procedure requires Jordan decomposition, which
introduces a lot of complications. In this article we will focus on how to compute defective MV function in real $\cl{p}{q}$. The article presents an extension of our previous article \cite{AcusMMAS2023} (which doesn't require neither explicit diagonalization nor Jordan decomposition procedure) to the case of defective MV or matrix. The described method can be used to verify our results of computation of MV functions in low dimensional ($n\le3$) Clifford algebras presented in \cite{AcusDargys2024log,AcusDargys2024}.

The matrix functions in general can be computed by a number of different ways
\cite{Sobczyk1997,Costa2004OnTE,Zhou2005,Fujii2007ExponentiationOC,Fujii2012,Herzig2014,Zela2014}. Our approach is based on the method
where renovation of spectral basis is employed, and which belongs to the class of {\it polynomial methods}~\cite{Sobczyk1997}.
In particular, the latter method is based on the minimal polynomial of considered MV. Then, instead of generalized spectral
decomposition procedure, which was used in~\cite{Sobczyk1997} to find explicit basis expansion coefficients, we provide recursive
formulas that greatly simplify the most problematic step in the computation. We believe that exact (closed form) formulas for exponentials and
other functions for low dimensional cases investigated in~\cite{Zela2014,Fujii2012,Fujii2007ExponentiationOC}, including matrices that are
representations of some Lie groups or have some special symmetries~\cite{Zhou2005,Costa2004OnTE}, now become relatively simple to
compute.

In Section~\ref{sec:charpoly} the methods to generate characteristic polynomials in \cl{p}{q} algebras characterized by
arbitrary signature $\{p,q\}$ are discussed. The minimal polynomial is introduced in Section~\ref{sec:minpoly}. In Section~\ref{sec:classical} we shortly remind classical method~\cite{Sobczyk1997},
and the Section~\ref{sec:main} presents main results of the article. Examples are given in Sectios~\ref{sec:exampleCl30}-\ref{sec:exampleImplicit} followed by Conclusion and Perspectives in Section~\ref{sec:discussion}.

\section{Notation and characteristic polynomial of MV}\label{sec:charpoly}

In GA, the geometric product of orthonormalized basis vectors
$\e{i}$ and $\e{j}$ satisfy~\cite{Lounesto1997} the
anti-commutation relation, $\e{i}\e{j}+\e{j}\e{i}=\pm
2\delta_{ij}$, where space denotes usual geometric product.  For a mixed signature \cl{p}{q} algebra, the
squares of basis vectors are $\e{i}^2=+1$ for first $p$ vectors,
$i=1,2,\ldots, p$, and $\e{j}^2=-1$ for remaining $q$ vectors,
$j=p+1,p+2,\ldots, p+q$. The sum $n=p+q$ is equal to dimension of
the vector space. The general MV is expressed as
\begin{equation}\label{mvA}
  \m{A}=a_0+\sum_{i}a_i\e{i}+\sum_{i<j}a_{ij}\e{ij}+\cdots+ a_{1\cdots n} \e{1\cdots n}
  =\sum_{J=0}^{2^n-1}a_J\e{J},
\end{equation}
where $a_i$, $a_{ij\cdots}$ are the real coefficients. The number
of subscripts in the basis element $\e{\cdots }$ indicates the grade
of the element. The ordered set of indices denoted by a single
capital letter $J$ is referred to as a multi-index. The basis
elements $\e{ij\cdots}$ are always assumed to be listed in the
inverse degree lexicographic order, i.e., the element with a lower grade is listed before
all elements of higher grades while lexicographically once the grade of both elements is the same.
For example, when $p+q=3$ the basis elements are listed in
the following order $\{1,\e{1},\e{2},\e{3},\e{12}\equiv\e{1}\e{2},\e{13}\equiv\e{1}\e{3},
\e{23}\equiv\e{2}\e{3},\e{123}\equiv\e{1}\e{2}\e{2}\equiv I\}$, i.e., both the number of indices and
their values always increases from left to right.

Characteristic polynomial $\chi_{\m{A}}(\lambda)$ of MV $\m{A}$ of variable $\lambda$
plays an important role in many applications. We will mainly use
it for general considerations and for comparison with minimal
polynomial of MV, which is described in next section. Every MV
$\m{A}\in\cl{p}{q}$ has a characteristic polynomial
$\chi_{\m{A}}(\lambda)$ of a degree $d$ in $\bbR$,
\begin{align}\label{CharPolyDef}
  \chi_{\m{A}}(\lambda)=\sum_{k=0}^{d} C_{(d-k)}(\m{A})\, \lambda ^k.
\end{align}
where
$d=2^{\lceil\tfrac{n}{2}\rceil}$ is the integer, $n=p+q$. In
particular, $d=2^{n/2}$ if $n$ is even and $d=2^{(n+1)/2}$ if $n$
is odd.

We will take\footnote{This choice ensures that the (symbolic)
coefficient $C_{(d)}(\m{A})$ (determinant of MV) remains syntactic positive
in any vector space of dimension $n$.} $C_{(0)}(\m{A})=-1$, then
$C_{(1)}(\m{A})$ will represent MV's trace:
$C_{(1)}(\m{A})=\Tr(\m{A})=d \left\langle\m{A}\right\rangle_{0}$,
where $\left\langle\m{A}\right\rangle_{0}$ is the scalar part of
MV in~\eqref{mvA}, i.e. $\left\langle\m{A}\right\rangle_{0}=a_0$.
The largest  coefficient $C_{(d)}(\m{A})$ gives MV determinant
with opposite sign, $C_{(d)}(\m{A})=-\Det(\m{A})$.

The characteristic polynomial can be computed in a number of ways.
We will use recursive Faddeev-LeVerrier-Souriau algorithm~\cite{Householder1975,Hou1998,Shirokov2021}, where each
recursion step produces one of the coefficients $C_{(k)}(\m{A})$
of the polynomial~\eqref{CharPolyDef}. The first recursion gives
$C_{(1)}(\m{A})$. Each subsequent  step produces the coefficient
$C_{(k)}(\m{A})=\frac{d}{k}\langle \m{A}_{(k)} \rangle_{0}$ and a
new MV $\m{A}_{(k+1)}=\m{A}
\bigl(\m{A}_{(k)}-C_{(k)}(\m{A})\bigr)$ according to
\begin{equation}\label{FLAlg}
  \begin{array}{rcl}
 \m{A}_{(1)}=\m{A}&\rightarrow&C_{(1)}(\m{A})=\frac{d}{1}\langle \m{A}_{(1)}\rangle_0,\\
 \m{A}_{(2)}=\m{A}\bigl(\m{A}_{(1)}-C_{(1)}(\m{A})\bigr)
&\rightarrow&C_{(2)}(\m{A})=\frac{d}{2}\langle \m{A}_{(2)}\rangle_0,\\
    &\vdots&\\
 \m{A}_{(d)}=\m{A}\bigl(\m{A}_{(d-1)}-C_{(d-1)}(\m{A})\bigr)
&\rightarrow&C_{(d)}(\m{A})=\frac{d}{d}\langle
\m{A}_{(d)}\rangle_0.
\end{array}
\end{equation}
The last step, as mentioned, returns the determinant with
opposite sign: $-\Det(\m{A})=\m{A}_{(d)}=C_{(d)}(\m{A})=\m{A}
\bigl(\m{A}_{(d-1)}-C_{(d-1)}(\m{A})\bigr)$. At $(d+1)$
step one gets the identity,
$\m{A}_{(d+1)}=\m{A}\bigl(\m{A}_{(d)}-C_{(d)}(\m{A})\bigr)=0$,
known as Cayley-Hamilton theorem. In particular, it states that if we replace
polynomial variable $\lambda$ by multivector $\m{A}$ in the
characteristic polynomial~\eqref{CharPolyDef} we still get
zero,
\begin{align}\label{CayleyHamiltonFormula}
  \sum_{k=0}^{d}  \m{A}^k C_{(d-k)}(\m{A})=\m{A}^{d}C_{(0)}(\m{A})+\m{A}^{d-1}C_{(1)}(\m{A})+\cdots + C_{(d)}(\m{A})= &0\,.
\end{align}
The above identity allows to write down multivector power
$\m{A}^d$ (and, of course, all higher powers) as a linear
combination of multivectors $\m{A}^0,\m{A},\m{A}^2,\ldots,\m{A}^{d-1}$, which correspond to respective matrices\footnote{Multivector representation by matrices is determined by Bott table~\cite{Lounesto1997}, p.~205, which defines isomorphism (understood as isomorphism among vector spaces, not algebras) between multivectors and matrices. The concepts of MV determinant and trace here are defined~\cite{Shirokov2021} without reference to the mentioned matrix representation.}.

The Cayley-Hamilton theorem also can be interpreted as a statement about linear dependency of vectors in higher dimensional space. Indeed, since geometric multiplication of MVs gives another MV of the same algebra, we can write all coefficients at (sorted) basis elements of MV in
a list and interpret them as the components of some vector in $2^n$ dimensional linear space. Alternatively, since $m\times m$ matrix
multiplication also yields a matrix of same dimension $m\times m$ we can reorder (flatten) $m\times m$ elements of a matrix into a column/row of length $m^2$ and also interpret them as vectors in some other linear space. Then Cayley-Hamilton theorem simply states that vectors constructed in both cases become linearly dependent if their number exceeds value $d$. We see, that in vector interpretation the role Cayley-Hamilton theorem plays (by setting maximal value $d$) is very similar to a role that vector space dimension $n$ plays by restricting 
maximal number of independent vectors in nD. Therefore, we can use the same universal algorithm to search for vector dependency. This will help us to understand the method of computation of the minimal polynomial of MV in next section, which is just a variant of search of independent vectors in given vector space.  

Let's now consider a Taylor series expansion\footnote{An attempt to compute MV function in this way, in general,  is a bad plan, since at first
sight even for a well behaved and convergent function some of the  coefficients may skyrocket to very high values before starting to decrease to
a true value. An example of bad behaviour can be found in paper~\cite{AcusDargys2024}.} of some MV function $f(\m{A})$ around a regular point of
the same function $f(x)$ of scalar argument, for example, around zero
\begin{align}\label{fTaylor}
  f(\m{A})=f(0)+ f^\prime(0)\m{A}+ \frac{1}{2!}f^{\prime\prime}(0)\m{A}^2+\cdots +
  \frac{1}{k!}f^{(k)}(0)\m{A}^{k}+\cdots.
\end{align}
The Cayley-Hamilton theorem ensures that all powers equal or large than $\m{A}^{d}$ can be expressed in terms of lower ones:
\begin{align}\label{fTaylorMaxd}
  f(\m{A})=\bigl(f(0)+\cdots\bigr) + \bigl(f^\prime(0)+\cdots\bigr)\m{A}+ \Bigl(\frac{1}{2!}f^{\prime\prime}(0)+\cdots\Bigr)\m{A}^2+\cdots +
  \Bigl(\frac{1}{(d-1)!}f^{(d-1)}(0)+\cdots\Bigr)\m{A}^{d-1},
\end{align}
where the coefficients of linear combination of all higher powers
of MV now have been  moved inside lower expansion coefficients
$\bigl(\frac{1}{k!}f^(k)(0)+\cdots\bigr)$  at $\m{A}^k,
k\le(d-1)$. The sum, nevertheless may be infinite. The question,
therefore, is ``can such a sum be summed up?''. It appears that one can
find a basis in which the sums at powers $\m{A}^k, k\le(d-1)$
become finite, i.e., we can explicitly do the summation. The basis will be called a generalized spectral basis~\cite{Sobczyk1997}.

\section{Minimal polynomial of MV}
\label{sec:minpoly}

In the matrix theory a most important polynomial is the so-called
minimal polynomial which we will denote $\mu_{\hat{A}}(\lambda)$. It establishes conditions for
diagonalizability of matrix $\hat{A}$. Matrices will be denoted by hats in order to distinguish them from MVs, which lack hats. The polynomial
$\mu_{\m{A}}(\lambda)$ may be defined for MV as well. It is well-known that matrix is diagonalizable (aka
non-defective) if and only if the minimal polynomial of the matrix
does not have multiple (repeated) roots, i.e., when the minimal
polynomial consists of product of distinct linear factors. We can use this condition as a definition and the criterion for diagonalizability (non-defectiveness) of MV too. Table~\ref{ChiMuComparison} compares characteristic and minimal
polynomial factors for trivial and non-trivial Jordan blocks in case of matrices $\hat{\m{A}}$.
\begin{table}
  \begin{center}
  \begin{tabular} {c|ccc}
    Jordan block &
    $\left.\begin{aligned}\begin{matrix}
      \lambda_i&0&\cdots&0\\
      0&\lambda_i&\cdots&0\\
      &&$\vdots$\\
0&&\cdots&\lambda_i
    \end{matrix}\end{aligned}\right\}k$ &
 $\left.\begin{aligned}\begin{matrix}
   \lambda_i&1&0&\cdots&0\\
      0&\lambda_i&1&\cdots&0\\
      &&&$\vdots$\\
      0&&\cdots&\lambda_i&1\\
   0&&\cdots&0&\lambda_i
    \end{matrix}\end{aligned}\right\}k$\\ \hline
    $\mu_{\hat{\m{A}}(x)}$ & $(x-\lambda_i)$ & $(x-\lambda_i)^k$ \\
    $\chi_{\hat{\m{A}}(x)}$  & $(x-\lambda_i)^k$ &  $(x-\lambda_i)^k$
\end{tabular}
\end{center}
  \caption{Minimal and characteristic polynomial factors in case of root $\lambda_i$ of multiplicity $k$ for
  trivial (left) and non-trivial (right) Jordan blocks of dimension $k$.}\label{ChiMuComparison}
\end{table}

In calculation of MV function $f(\m{A})$ a central role is
played by MV minimal polynomial $\mu_{\m{A}}(\lambda)$. For MV $\m{A}\in\cl{p}{q}$ it is a
polynomial in variable $\lambda$ of degree $d\le
2^{\lceil\tfrac{n}{2}\rceil}$,
\begin{equation}\label{MinPolyDef}
  \mu_{\m{A}}(\lambda)=\mu(\lambda)=\sum_{k=0}^{d} C_{(d-k)}(\m{A})\, \lambda^k,
\end{equation}
where the coefficients\footnote{We intentionally use the same notation for coefficients both for characteristic and
minimal polynomials, since  for diagonalizable MV coefficients of both polynomials coincide
(in the article they have opposite signs for computational purposes)
and we will deal  further with the minimal polynomials only.} $C_{(d-k)}(\m{A})$ characterize  MV/matrix,
or a general linear operator. We will take the coefficient at
$\lambda ^d$ to be normalized to unity,
 i.e. $C_{(0)}(\m{A})=1$ (for characteristic polynomial we used $C_{(0)}(\m{A})=-1$). It
is well-known~\cite{Garibaldi2004} that the minimal polynomial $\mu_{\m{A}}(\lambda)$
divides the characteristic polynomial $\chi_{\m{A}}(\lambda)$,
therefore, in general the former is of the same or lower degree than later,
Table~\ref{ChiMuComparison}. If all roots of characteristic
equation $\chi_{\m{A}}(\lambda)=0$ are different, in other word
the roots have  multiplicity $1$, then $\chi_{\m{A}}(\lambda)$ and
$\mu_{\m{A}}(\lambda)$ may differ by constant factor (normalization coefficient) only.

To have shorter notation, instead of $\mu_{\m{A}}(\lambda)$ we
shall write $\mu$, since $\mu$ depends on MV $\m{A}$ indirectly
through coefficients $C_k$. We will also introduce an alternative
notation with superscript, $\mu(\lambda)=\mu^{(0)}(\lambda)$, in
order to make the notation compatible with that for $k$-th
derivative of minimal polynomial,
$\mu^{(k)}(\lambda)=\frac{1}{k!}\frac{\dd^k \mu(\lambda)}{\dd
\lambda^k}$.  Taking into account that the factorial of zero is
$0!=1$, the notation $\mu^{(0)}(\lambda)$ simply means a ``zero''
derivative of minimal polynomial, i.e.
$\mu^{(0)}(\lambda)\equiv\mu(\lambda)$. The first derivative of
$\mu(\lambda)$ then is $\mu^{(1)}(\lambda)=\frac{1}{1!}\frac{\dd
\mu(\lambda)}{\dd \lambda}=\sum_{k=1}^{d} k\,
C_{(d-k)}(\m{A})\lambda ^{k-1}=\sum_{k=0}^{d-1} (k+1)
C_{(d-k-1)}(\m{A})\lambda ^{k}$,
where in the last sum the
summation index was shifted to start from zero in order to be consist with
the formula (16) in paper~\cite{AcusMMAS2023} (see factor $\beta$).

The algorithm~\ref{algMinimalPoly} presented below explains in terms
of GA  how the MV minimal polynomial can be programmed.\footnote{The
algorithm in the paper~\cite{AcusMMAS2023} didn't take into account situation when number of vectorList elements exceed the  degree of characteristic polynomial. In the latter case, one must reduce the degree of polynomial by one. This minor correction is now represented by {\bf
If} statement in the Algorithm~\ref{algMinimalPoly}.}

\begin{algorithm}
    \caption{Minimal polynomial of MV}\label{algMinimalPoly}
    \begin{algorithmic}[1] 
      \Procedure{MinimalPoly}{$\m{A},x$} \Comment{Input is the MV $\m{A}=\sum_{J=0}^{2^n-1}a_J\e{J}$ and polynomial variable $x$}
      \State nullSpace=\{\};\quad lastProduct=1;\quad vectorList=\{\}; \Comment{Initialization; \{\} is an empty list}
            \While{nullSpace===\{\}} \Comment{Keep adding MV coefficient vectors to vectorList until the null space gets to be nonempty}
                \State lastProduct$\gets\m{A}\circ$lastProduct\;
                \State  vectorList$\gets$AppendTo[vectorList,\, toCoefficientList[lastProduct]]\;
                \State  nullSpace$\gets$NullSpace[Transpose[vectorList]];
            \EndWhile\label{nullSpaceEnd}\Comment{Construct minimal polynomial from nullSpace coefficients and powers of input variable $x$}
      \State     \If{Length[nullSpace] $\le$ $\mathrm{deg}(\chi)$}  \Comment{If length exceeds the degree of characteristic polynomial, reduce degree by one}
    \State \textbf{return} $\mathrm{First[nullSpace]}\cdot \{x^1,x^2,\ldots,x^{\mathrm{Length[nullSpace]}}\}$
 \Else
    \State \textbf{return} $\mathrm{First[nullSpace]}\cdot \{x^0,x^1,x^2,\ldots,x^{\mathrm{Length[nullSpace]-1}}\}$
  \EndIf

        \EndProcedure
    \end{algorithmic}
\end{algorithm}
The algorithm was explained earlier in paper~\cite{AcusMMAS2023}, else the computation closely follows the algorithm for a matrix minimal
polynomial in~\cite{Mathworld2022}. Because in GA the orthonormal basis elements $\e{J}$ are linearly independent by definition, we can
construct vectors from MV coefficients, in an exactly the same way as it was described in the interpretation of Cayley-Hamilton
formula~\eqref{CayleyHamiltonFormula}. The Algorithm~\ref{algMinimalPoly} performs a search until the vectors constructed from
coefficients at powers of MV become linearly dependent. Once this has been achieved\footnote{This is guaranteed by the
Cayley-Hamilton theorem which claims that powers of $\m{A}^d$ or greater are linear combinations of lower powers of $\m{A}$.}, the {\bf NullSpace[~]} returns a list of coefficients of the linear combination. These coefficients then are multiplied by corresponding power of $\m{A}$  and summed up to form a minimal MV polynomial.

Since output of {\it Mathematica} command {\bf NullSpace[~]} is a bit specific it deserves more detailed explanation. Suppose we want to compute the minimal polynomial of the idempotent  $\m{a}=\frac12 (1 + \e{1})$ of $\cl{3}{0}$ algebra with orthonormal basis listed in section~\ref{sec:charpoly}. Following the Algorithm~\ref{algMinimalPoly} we start by forming list of components of $\m{a}, \m{a}^2, \m{a}^3, \ldots$. A first vector  $\m{a}$ added to nullSpace variable has components $\{1/2,1/2,0,0,0,0,0,0\}$. In the second step we add vector components $\{1/2,1/2,0,0,0,0,0,0\}$ of $\m{a}^2$.  Since $\m{a}$ is an idempotent, i.e. $\m{a}^2=\m{a}$, the components of  $\m{a}^2$ are exactly the same. The variable nullSpace  now contains two identical vectors $\textrm{nullSpace}=\{\{1/2,1/2,0,0,0,0,0,0\},\{1/2,1/2,0,0,0,0,0,0\}\}$, which are, obviously, linearly dependent. The {\bf NullSpace[~]} returns $\{-1,1\}$, which simply states that $-\m{a}+\m{a}^2=0$. The computed minimal polynomial of $\m{a}$, therefore is $-x+x^2$, which coincides with the minimal polynomial of matrix representation of $\m{a}$ by complex $2\times 2$ Pauli like matrices
$\hat{\e{}}_1= \left(
\begin{smallmatrix}
 1 & 0 \\
 0 & -1 \\
\end{smallmatrix}
\right)$, $\hat{\e{}}_2= \left(
\begin{smallmatrix}
 0 & 1 \\
 1 & 0 \\
\end{smallmatrix}
\right)$, $\hat{\e{}}_3=\left(
\begin{smallmatrix}
 0 & -\ii \\
 \ii & 0 \\
\end{smallmatrix}
\right)
$.
It should be noted, however, that this is not always the case. Indeed, since real $\cl{3}{0}$ Clifford algebra is represented by complex $2\times 2$ matrices, one can easily find an example when the minimal polynomial of MV and the minimal polynomial it's matrix representation are distinct. For example, if we compute a minimal polynomial of an algebraic spinor $\psi=\frac{1}{2} c_1 (\e{1}+1)+\frac{1}{2} c_3 (\e{2}-\e{12})+\frac{1}{2} c_2 (\e{123}+\e{23})+\frac{1}{2} c_4 (\e{13}-\e{3})$, which was computed using the idempotent $\m{a}$ above, we get $\mu_\psi(x)=x^3-2 c_1 x^2+\left(c_1{}^2+c_2{}^2\right) x$. Obviously it can't be a minimal polynomial of the matrix representation, which is $2\times 2$ complex matrix (and, therefore, can't be of degree $3$). Indeed, the matrix representation of the above spinor is $\hat{\psi}=\left(
\begin{smallmatrix}
 c_1+i c_2 & 0 \\
 c_3-i c_4 & 0 \\
\end{smallmatrix}
\right)$. The minimal polynomial $\mu_{\hat{\psi}}(x)=x^2+(-c_1-\ii c_2) x\,$ of this matrix is of degree $2$. Note, that in contrast to real polynomial $\mu_\psi(x)$ of MV, the polynomial $\mu_{\hat{\psi}}(x)$ of the matrix representation has complex coefficients and is not suited for real Clifford algebra $\cl{3}{0}$, since we know that geometric multiplication of MVs with real coefficients can produce real expressions only. The same remark applies for characteristic polynomial in GA, which is now well understood~\cite{Shirokov2021}.

The described method of computation of minimal polynomial must bring in a  clear difference
between the characteristic and minimal polynomials: The characteristic polynomial is able to provide general condition when generic vectors
must become linearly dependent only, whereas the minimal polynomial finds strict linear relation between particular vectors. For example, while
randomly generated 3~vectors (with integer coefficients) in 3D~space are linearly independent in most cases, it may happen that in
rare cases they all lie in the same~2D plane, or even be collinear. The characteristic polynomial can't distinguish between these cases, while
minimal polynomial sensitively represents each case.

At the end we would like to comment briefly on computation of minimal polynomial in {\it Mathematica} language syntax (also refer to
Algorithm~\ref{algMinimalPoly}). The full implementation is provided in GA package~\cite{AcusDargys2025}. The command-functions that begin with a capital letter, {\bf AppendTo[\,], NullSpace[\,], Transpose[\,], Last[\,], First[\,]}, and  {\bf Dot[\,]} (short form of which is $\,\cdot\,$),
are internal {\it Mathematica} commands. Exceptions are the symbol $\circ$ (geometric product) and {\bf toCoefficientList[\,]} transformation.
The latter is very simple. It takes a multivector $\m{A}$ and constructs a vector (list) from its coefficients:
{\bf toCoefficientList[$a_0+a_1\e{1}+a_2\e{2}+\cdots+a_I I$]}$\to \{a_0,a_1,a_2,\ldots,a_I \}$. A real job is done by {\it Mathematica}  function {\bf NullSpace[\,]}, which searches for linear dependency of the incrementally augmented vector list {\bf vectorList} (see Algorithm~\ref{algMinimalPoly}). After such a list of vectors has been found (this is guaranteed by Cayley-Hamilton theorem) the function outputs the linear combination for which the sum of vectors turns to zero. The command {\bf AppendTo[vectorList, newVector]} appends the {\bf newVector} to the list {\bf vectorList} of already checked vectors.

\section{Calculation of MV function by classical method}
\label{sec:classical}

To help the reader grasp the main result presented in the next
section, here  we will review the classical method how to find  a
generalized spectral basis and then apply it to compute the function of multivector, matrix or 
operator~\cite{Sobczyk1997}.

Construction of the basis starts from minimal polynomial which
in a factorized form can be written as
\begin{align}\label{cl:minpolyfactored}
  \mu(x) =\prod_{i=1}^r(x-\lambda_i)^{m_i}\,.
\end{align}
Here $r$ stands for different eigenvalues $\lambda_i$ of
multiplicity $m_i$. The degree of the minimal polynomial $\mu(x)$ is
$\sum_{i=1}^r m_i \le 2^{\lceil\tfrac{n}{2}\rceil}$. For
convenience, the roots are listed in increasing multiplicity order:
$1\le m_1\le m_2\le\cdots\le m_r$. So, the last root has the
largest multiplicity. The generalized spectral decomposition of
the MV begins by decomposing the  inverse of a minimal polynomial
into partial fractions,
\begin{align}\label{cl:minpolyPFD}
  \frac{1}{\mu(x)} =&\sum_{i=1}^r \biggl(\frac{a_0}{(x-\lambda_i)^{m_i}}+\frac{a_1}{(x-\lambda_i)^{m_i-1}}+\cdots + \frac{a_{m_i-1}}{(x-\lambda_i)} \biggr)\notag \\
  =&\frac{h_1(x)}{(x-\lambda_1)^{m_1}}+\cdots + \frac{h_{r}(x)}{(x-\lambda_r)^{m_r}}\notag \\
   =&\frac{h_1(x)\psi_1(x)+\cdots + h_r(x)\psi_r(x)}{\mu(x)}\,,
\end{align}
where $h_i(x)=\sum_{s=0}^{m_i-1} a_s (x-\lambda_i)^s$ and $\psi_i(x)=\prod_{j\neq i}^r(x-\lambda_j)^{m_j}$, i.e. we have divided each
term of the sum into part $h_i(x)$ that includes the root $\lambda_i$ and the part $\psi_i(x)$ that is free of the root $\lambda_i$. After
comparison with the initial expression we see that $\sum_{i=1}^r h_i(x)\psi_i(x)= h_1(x)\psi_1(x)+\cdots + h_r(x)\psi_r(x)=1$. Therefore, we can
define a polynomials $p_i(t)= h_i(t)\psi_i(t)$ having the property $\sum_{i=1}^r p_i=1$. Since  $\psi_i(t)\psi_j(t)$ is divisible\footnote{All polynomial products should be understood as products modulo $\mu(x)$. For more details, how these properties can be verified see
example in Section~\ref{sec:exampleCl30}.} by  $\mu(t)$ for all $i\neq j$, we also have $p_i p_j=0$. Thus, we conclude that $p_i^2=(p_1+\cdots
+p_r)p_i=1\, p_i=p_i$. The obtained properties imply that the operators $p_i$ make a set of mutually annihilating idempotents that realize
partition of the unity.

Now, for each $1\le i\le r$ let's define a new polynomial $q_i=(x-\lambda_i)p_i$. For non-repeating root $m_i=1$ we see that $q_i=0$.
For repeated  root of multiplicity $m_i>1$ we have $q_i^{m_i-1}\neq 0$, but $q_i^{m_i}= 0$. Thus a list $q_1, q_2,\cdots,q_r$ is a set of
nilpotents with corresponding nilpotency indices $m_1,m_2,\cdots,m_r$. If $i\neq j$, we have $p_j q_i=0$. Therefore, $p_iq_i=(p_1+\cdots
+p_r)q_i=1\, q_i=q_i$, and one says that nilpotents $q_1, q_2,\cdots,q_r$ are projectively related to idempotents $p_1, p_2,\cdots,p_r$. As a
result we have got the following generalized spectral basis
\begin{align}\label{genSpectralBasis}
\bigl\{
  \underbrace{p_1,q_1,q_1^{2}\cdots q_1^{m_1-1}}_{m_1}  \cdots \underbrace{p_i,q_i,q_i^{2}\cdots q_i^{m_i-1}}_{m_i},\cdots
  \underbrace{p_r,q_r,q_r^{2}\cdots q_r^{m_r-1}}_{m_r}
\bigr\},
\end{align}
where $p_i=p_i(x),q_i=q_i(x)$ are polynomials.
If the multiplicity $m_i$ of root $\lambda_i$ is 1, the corresponding basis
blocks (indicated by underbraces) consist of a single polynomial
$p_i$.

\begin{lemma}
The generalized spectral decomposition of linear operator can be
written as
\begin{align}\label{genSpectralDecomposition}
\m{A}=\sum_{i=1}^r \bigl(\lambda_i +q_i\bigr)p_i\,.
\end{align}
  We need to show that $\m{A} p_i = (\lambda_i+q_i) p_i$.
  Since $q_i = (\m{A}-\lambda_i) p_i$,
  $\m{A} p_i = q_i + \lambda_i p_i = q_i p_i + \lambda_i p_i$, where we used the property $p_iq_i=q_i$.  Because $p_i$ is the idempontent, $p_i^2=p_i$, after expansion we have $\m{A}=\m{A}\sum_{i=1}^r p_i=\m{A}$.
\end{lemma}

Now higher powers of $\m{A}$  can be easily computed:
\begin{alignat}{5}\label{genSpectralDecompositionPowers}
  & 1\quad&=&\quad p_1\quad&+&\quad p_2\quad &+&\,\cdots\, &+&\quad p_r\notag\\
  & \m{A}\quad&=&\quad (\lambda_1+q_1)p_1\quad&+&\quad  (\lambda_2+q_2)p_2\quad  &+&\,\cdots\, &+&\quad  (\lambda_r+q_r)p_r\notag\\
& \m{A}^2\quad&=&\quad (\lambda_1+q_1)^2 p_1\quad&+&\quad  (\lambda_2+q_2)^2 p_2\quad  &+&\,\cdots\, &+&\quad  (\lambda_r+q_r)^2 p_r\\
  &\vdots&&&&&&&&\notag\\
  & \m{A}^{m-1}\quad&=&\quad (\lambda_1+q_1)^{m-1}p_1\quad&+&\quad  (\lambda_2+q_2)^{m-1} p_2\quad  &+&\,\cdots\, &+&\quad  (\lambda_r+q_r)^{m-1} p_r\, .\notag
\end{alignat}
To find $(\lambda_j+q_j)^{k}$ one can use the  binomial theorem and  property $q_i^{m_i}=0$,
\begin{align}\label{binomialF}
  (\lambda_j+q_j)^{k}=\lambda_j^k+\binom{k}{1}\lambda_j^{k-1}q_j+\cdots+\binom{k}{m_j-1}\lambda_j^{k-(m_j-1)}q_j^{m_j-1}\,.
\end{align}
Here $\binom{k}{p}=\frac{k!}{p!(k-p)!}$ denotes the usual binomial coefficient for $p\le k$ and zero otherwise.

Once we know how to compute the powers of MV, we can apply this property to function series in MV argument (we suppose the function is
analytic). For simplicity, let's  assume that the function of a scalar (complex) argument, $g(z)$, has a convergent Taylor series in the
neighborhood of the eigenvalue $\lambda_i$. Then, MV function  $g(\m{A})$ can be defined as Taylor series with  scalar variable $z$
replaced by MV\footnote{The Cayley-Hamilton theorem, of course, ensures that in the sum there exists a maximal power of $\m{A}^d$. Nonetheless,
in arbitrary basis the series remains infinite because each $\m{A}^k$ power, $k>d$, that exceed dimension $d$ will contribute to all
coefficients of lower powers of $\m{A}$. Now we see, the main reason why the generalized spectral basis is so useful is that the coefficient
sums terminate in this basis at $k\le d$.} $\m{A}$, i.e., $g(\m{A})=\sum_{k=0}^\infty \m{A}^k = a_0+ a_1 \m{A}+a_2\m{A}^2+\cdots$.
Using~\eqref{genSpectralDecompositionPowers} and properties $p_i^2=p_i$, $q_i^{m_i}=0$, $p_iq_i=q_i$, $p_j q_i=0$, we find,
\begin{align}\label{gTaylor}
  g(\m{A})=&\phantom{+}\,a_0\bigl( p_1+ p_2 +\cdots + p_r\bigr)\notag \\
  &+a_1\bigl((\lambda_1+q_1) p_1+ (\lambda_2+q_2)p_2 +\cdots + (\lambda_r+q_r)p_r\bigr)\notag \\
  &+a_2\bigl((\lambda_1+q_1)^2 p_1+ (\lambda_2+q_2)^2 p_2 +\cdots + (\lambda_r+q_r)^2 p_r\bigr)\notag \\
  &+a_3\bigl((\lambda_1+q_1)^3 p_1+ (\lambda_2+q_2)^3 p_2 +\cdots + (\lambda_r+q_r)^3 p_r\bigr)+\cdots\notag \\
  =&\bigl(a_0+a_1\lambda_1+a_2\lambda_1^2 + a_3\lambda_1^3+\cdots\bigr)\notag\\
  &+\left(a_1+\binom{2}{1}a_2\lambda_1+\binom{3}{1}a_3\lambda_1^2 +\cdots \right)q_1\notag\\
  &+\left(a_2+\binom{3}{2}a_3\lambda_1+\binom{4}{2}a_4\lambda_1^2 +\cdots \right)q_1^2+\cdots+\text{sums for rest $r-1$ roots}\\
  =&g(\lambda_1)p_1+ g^\prime(\lambda_1)q_1+ \frac{1}{2!}g^{\prime\prime}(\lambda_1)q_1^2+\cdots +
  \frac{1}{(m_1-1)!}g^{(m_1-1)}(\lambda_1)q_1^{(m_1-1)}+\cdots\notag\\
  &+g(\lambda_r)p_r+ g^\prime(\lambda_r)q_r+ \frac{1}{2!}g^{\prime\prime}(\lambda_r)q_r^2+\cdots +
  \frac{1}{(m_1-1)!}g^{(m_1-1)}(\lambda_r)q_r^{(m_r-1)}\, ,\notag
\end{align}
where the primes denote derivatives, $g^\prime(\lambda_1)\equiv g^{(1)}(\lambda_1)=\frac{\dd g(z)}{\dd z}\vert_{z=\lambda_1}$ etc.
Introduction of notation
\begin{align}\label{gNotation}
  g(\lambda_i+q_i)=g(\lambda_i)p_i+ g^\prime(\lambda_i)q_i+ \frac{1}{2!}g^{\prime\prime}(\lambda_i)q_i^2+\cdots +
  \frac{1}{(m_i-1)!}g^{(m_i-1)}(\lambda_i)q_i^{(m_i-1)},
\end{align}
then allows compactly write down the above finite sum as
\begin{align}\label{gTaylorFin}
  g(\m{A})=\sum_{i=1}^r g(\lambda_i+q_i)p_i \,.
\end{align}

The classical formula~\eqref{gTaylorFin} may be used to compute
analytical functions of MV, matrices or general linear
operators. Note that almost all steps of symbolic manipulations (the
decomposition of inverse of minimal polynomial, the
Taylor expansion) can be done using formal scalar
argument. The essential and most tricky part of these computations
is, of course, the decomposition of inverse of minimal polynomial
$1/ \mu(x)$, which makes a core of all computations. The above
description (intentionally) lacks any details on this subtle point.
The only important feature of classical computation method we want to stress is
the order in which polynomials $p_i, q_i^k$ are computed. In particular, the polynomials $p_i$ are computed at first.
Only then it is possible to compute powers of $q_i, q_i^2, \ldots$ in increasing order.
Because of latter property, the classical method requires operation of division (reduction) of polynomials.
This is a difficult task when minimal polynomial happens to be an irreducible
polynomial of high degree. In the article~\cite{Sobczyk1997} the decomposition is
implemented by using {\it Mathematica}'s universal command {\bf Series[~]}
and the fact that the expansion series must terminate after $m_i$ terms (the required program code is listed in
the appendix of~\cite{Sobczyk1997}). However the {\bf Series[~]} command uses general state-of-the-art
algorithms, where a short note on implementation of the command\footnote{See
\url{https://reference.wolfram.com/language/tutorial/SomeNotesOnInternalImplementation.html}.}
states that it ``works by recursively composing series expansions of
functions with series expansions of their arguments''. In the next
section we will show how to realize this decomposition explicitly,
without any reference to  command {\bf Series[~]}.

\section{MV function in \cl{\lowercase{p}}{\lowercase{q}} algebra}
\label{sec:main}
This section presents main result of the article and extends our earlier
efforts~\cite{AcusMMAS2023} to the case of non-diagonalizable MVs in a basis-free form.

\begin{proposition}[MV function in a basis-free form]\label{theoremMain}
The function $f(\m{A})$ of multivector argument
$\m{A}=\sum_{J=0}^{2^n-1}a_J\e{J}$ in $\cl{p}{q}$ algebra is the
MV that can be computed by the following explicit formula
\begin{align}\label{funcCoordFree}
f(\m{A})=
  & \sum_{i=1}^{r}\sum_{t=0}^{m_i-1} \Bigl(\frac{\dd^{t} f(\lambda_i+\tau)}{\dd \tau^t}\big\vert_{\tau=0}\Bigr) Q^{m_i-1-t}_i(\m{A},\lambda_i),
\end{align}
  where $m_i$ denotes multiplicity of root $\lambda_i$ (with the index $i$) of the minimal polynomial $\mu(\lambda)$ of the MV $\m{A}$. The polynomial $\mu(\lambda)$ has $r$ different roots. The symbol $\frac{\dd^{t} f(\lambda_i+\tau)}{\dd \tau^t}\big\vert_{\tau=0}$
  denotes the $t$-th derivative of function $f$ at the root $\lambda_i$ (zero derivative of a function is understood as the function itself).
  The MV polynomials\footnote{Since $\m{A}$  appears in $Q^{m_i-1-t}_i(\m{A},\lambda_i)$ explicitly,
  the latter is a polynomial both in $\lambda_i$ and $\m{A}$ (of course, the coefficients $C_{(k)}$ still are
  functions of $\m{A}$ as well). The same notation applies to polynomials $S^{(k)}(\m{A},\lambda)$ below.}
  $Q^{m_i-1-t}_i(\m{A},\lambda_i)=Q_i^{m_i-1-t}(\m{A},\lambda)\vert_{\lambda=\lambda_i}$ are recursively computed as
  follows
  \begin{align}\label{recursionQ}
  Q_i^{m_i-1}(\m{A},\lambda)=&\frac{S^{(0)}(\m{A},\lambda)}{\mu^{(m_i)}(\lambda)}
    \notag\\
  Q_i^{m_i-2}(\m{A},\lambda)=&\frac{S^{(1)}(\m{A},\lambda)- Q_i^{m_i-1}(\m{A},\lambda)\mu^{(m_i+1)}(\lambda)}{\mu^{(m_i)}(\lambda)}
  \notag\\
  Q_i^{m_i-3}(\m{A},\lambda)=&\frac{S^{(2)}(\m{A},\lambda)- Q_i^{m_i-2}(\m{A},\lambda)\mu^{(m_i+1)}(\lambda)- Q_i^{m_i-1}(\m{A},\lambda)\mu^{(m_i+2)}(\lambda)}{\mu^{(m_i)}(\lambda)}
  \notag\\ \vdots\notag\\
  Q_i^{0}(\m{A},\lambda)=&\frac{S^{(m_i-1)}(\m{A},\lambda)- Q_i^{1}(\m{A},\lambda)\mu^{(m_i+1)}(\lambda)- Q_i^{2}(\m{A},\lambda)\mu^{(m_i+2)}(\lambda)-\cdots -Q_i^{m_i-1}(\m{A},\lambda)\mu^{(2m_i-1)}(\lambda)}{\mu^{(m_i)}(\lambda)}
  \,.
\end{align}
The quantities $\mu^{(k)}(\lambda)$ are weighted $k$-th
  derivatives of minimal polynomial $\mu^{(k)}(\lambda)=\frac{1}{k!}\frac{\dd^k \mu(\lambda)}{\dd
  \lambda^k}$ (defined in Section~\ref{sec:minpoly}). The polynomials $S(\m{A},\lambda)=S^{(0)}(\m{A},\lambda)$
and and their weighted $k$-th derivatives $S^{(k)}(\m{A},\lambda)$ are defined by
  \begin{align}\label{formulasS}
  &S(\m{A},\lambda)=S^{(0)}(\m{A},\lambda)=\sum_{k=0}^{r-1} \biggl(\sum_{s=0}^{r-k-1}\lambda^s C_{(r-s-k-1)}(\m{A})\biggr)\, \m{A}^{k}, &&S^{(0)}(\m{A},\lambda_i)=S^{(0)}(\m{A},\lambda)\Big\vert_{\lambda=\lambda_i}\notag \\
  &\phantom{S(\m{A},\lambda)=}S^{(k)}(\m{A},\lambda)=\frac{1}{k!}\frac{\dd^k S(\m{A},\lambda)}{\dd \lambda^k},&&S^{(k)}(\m{A},\lambda_i)=S^{(k)}(\m{A},\lambda)\Big\vert_{\lambda=\lambda_i}\, .
\end{align}
\end{proposition}
\begin{proof}
At the moment we have proved formulas in the Proposition~\ref{theoremMain} for Clifford algebras with $p+q=4$ and
  $p+q=6$ by symbolically verifying them for defective MVs with all possible combinations of nontrivial Jordan blocks of algebras that have real matrix representations\footnote{At the moment we don't know an easy way to construct defective MV systematically for algebra that has complex and especially quaternion matrix representation. Nevertheless, since we know how to compute minimal polynomial for randomly generated MV, we can check the formula for arbitrary algebras, albeit not so systematically.}.  Also, we have checked formulas numerically (using exact arithmetic) for $p+q\le 12$, for this purpose using MVs with integer
  coefficients. Since the proposition is formulated in terms of polynomials with unspecified coefficients,
  a general proof could be tedious and will require manipulation/summation of products of coefficients of these polynomials.
  This task is postponed for further research.
\end{proof}

\begin{remark}[Generalization]
  It is straightforward to generalize the Proposition~\ref{theoremMain}  for arbitrary (finite dimensional)
  square matrices and linear operators. Indeed, in order to compute a function of matrix it is enough to replace
  MV geometric product by matrix product $\m{A}^k\to \underbrace{\hat{\m{A}}\cdot \hat{\m{A}}\cdots \hat{\m{A}}}_{k~\text{terms}}$.
  Computation of minimal polynomial of finite matrices is a well-known procedure.
\end{remark}

\begin{remark}[Recursion formula analogy]
 The recursion formulas~\eqref{recursionQ} reminds one of the procedure of construction of
  basis vectors for some irreducible representation~\cite{Georgi2000} of Lie group.
 Since we already know from Section~\ref{sec:classical} that $Q^{m_i}=0$
 the construction starts from a single basis vector of the highest weight $Q^{m_i-1}$.
 Then one defines a lowering operator,  application of which in succession produces basis vectors
 by one lower weight in each step. The procedure is repeated until the lowest weight basis vector of the
 representation is reached\footnote{We have chosen a case which most closely matches our algorithm.
 Sometimes the irreducible representation basis is computed starting from the lowest weight vector and then
 applying raising operator}. One could observe
 that the polynomial $Q_i^{m_i-1}$ corresponds to the polynomial $q_i^{m_i-1}$ of the highest
 power in a classical method.
 Indeed  (see Section~\ref{sec:classical}, paragraph above formula~\eqref{genSpectralBasis}), multiplication of $q_i^{m_i-1}$
 by $q_i$ results in zero, $q_i\,q_i^{m_i-1}=q_i^{m_i}=0$. Therefore $Q_i^{m_i-1}$ is exactly the polynomial from which
 the entire recursive lowering procedure starts. After subtraction of all previous $Q_i^k$ contributions in the last step
 we are left with the polynomial $Q_i^{0}(\m{A},\lambda)$ which has been denoted by $p_i$ in the classical method.
 Since explicit expressions for $Q_i^{k}(\m{A},\lambda)$ at $k>0$ coincide with those at $k\le 0$, they
 have exactly the same properties
 as listed in Section~\ref{sec:classical}  and, of course, the properties of $Q_i^{0}(\m{A},\lambda)$ are identical to $p_i$.
 Therefore, the recursion formulas~\eqref{recursionQ}, in fact, implement partial fraction decomposition
 of inverse minimal polynomial in a reverse order, what makes polynomial reduction redundant.
\end{remark}

\begin{remark}[Denominators]
  One should observe that denominators of~\eqref{recursionQ} enclose the weighted $m_i$-th derivative
  in the  minimal polynomial $\mu^{(m_i)}(\lambda)$. In fact, it prevents the denominator at root $\lambda_i$ of
  multiplicity $m_i$ to vanish.
  Indeed at the root $\lambda_i$ of multiplicity $m_i$ we have
  $\mu_{\m{A}}(x)=(x-\lambda_i)^m_i p(x)$, where polynomial $p(x)$ encompasses contribution from
  all other roots $\lambda_k\neq\lambda_i$.  After differentiation $m_i$ times we get an expression
  in which $(x-\lambda_i)$ disappears. Since $p(\lambda_i)\neq 0$, the denominator does not reduce to zero
  at $x=\lambda_i$. This explains why in case of a non repeating roots in equation~(16)
  in paper~\cite{AcusMMAS2023} the expression for $\beta$ factor in the denominator has only the  first derivative of
  minimal polynomial.
\end{remark}

\begin{remark}[Number of derivatives of MV function]
  From~Eq.\eqref{funcCoordFree} one can see that the non-diagonalizable MV function demands computation of the function
 derivatives of order that is by one less than the maximal multiplicity of a set of roots. From this  follows that MV function
 that has no repeated roots, i.e. the MV is non-defective, does not require computation of function derivatives at all.
 This is in agreement with formula~(16) in paper~\cite{AcusMMAS2023}.
\end{remark}

\begin{remark}[Evaluation]
Symbolic recursion formulas are same\footnote{Even if multiplicities of roots are distinct the symbolic expression has a lot of repeated parts, and computations can still
be highly optimized, see example in Sections~\ref{sec:exampleCl42}.} for every root $\lambda_i$ of the same multiplicity,  and consequently the
expressions for them are identical up until a particular root value will be inserted. Therefore, it is more efficient at first to do
all computations with a formal root $\lambda$ symbolically and then later to substitute a particular root values in a final step of
  computation. Moreover, we can compute entire generalised spectral basis using dummy commutative variable $x$ and make replacements $x^k\to \m{A}^k$ in a very last step of computations (in a matrix case the replacement reads $x^k\to \underbrace{\hat{\m{A}}\cdot \hat{\m{A}}\cdots \hat{\m{A}}}_{k~\text{terms}}$).
\end{remark}

\begin{remark}[Characteristic polynomial]
  We found it some how surprising that instead of the minimal polynomial $\mu(\m{A})$ in \eqref{recursionQ} we can use the characteristic polynomial $\chi(\m{A})$ or any other polynomial $p(\m{A})$, which yields zero for a given MV $\m{A}$. Indeed, if $S^{(0)}(\m{A},\lambda_i)$ is zero, then $Q_i^{m_i-1}(\m{A},\lambda_i)$ is zero as well. It means that $Q_i^{m_i-2}(\m{A},\lambda_i)$ becomes the polynomial $Q_i^{m_i^\prime-1}(\m{A},\lambda_i)$, with shifted index $i^\prime=i-1$ (since a differentiation $S^{(1)}(\m{A},\lambda_i)$ reduces polynomial degree by one). Therefore, an arbitrary polynomial what satisfy $p(\m{A})=0$ can be used. Of course, computations then becomes less effective, since in general we need to compute higher powers of MV for substitution in a final expression. It also suggest, that in principle, only eigenvalues and estimation of largest multiplicity is required to start the whole computation.
\end{remark}

Since the roots of a characteristic equation in real GA  in general are
complex numbers, the individual terms in the sums in Proposition~\ref{theoremMain}, strictly
speaking, are complex. However, the coefficients at basis elements
in the final result for some functions (for example, for exponential function) will simplify to real numbers. The real expressions for coefficients (when possible) can be obtained by summing complex conjugate roots pairwise to real expressions exactly in the same way as it was explained  in~\cite{AcusMMAS2023}.

\section{Example: defective MV in  \cl{3}{0}}\label{sec:exampleCl30}

The section  demonstrates how to apply Proposition~\ref{theoremMain} to compute arbitrary GA function of a non-diagonalizable
MV, i.e. of MV minimal polynomial of which has at least one repeated root.
In the examples below we will compute the exponential function of MV, since the obtained answer
then can be easily checked by using the defining property of the exponent, namely,
\begin{align}
\m{A}\exp(\m{A})=\left.\frac{\partial\exp(\m{A}t)}{\partial t}\right|_{t=1}\,.\label{expproperty}
\end{align}
To find arbitrary function of MV it is enough to replace the exponent and its scalar derivatives by corresponding function and
its derivatives in the examples presented below. The only restriction is that the function (and the required derivatives) should be well defined
at the roots of MV minimal polynomial.

\begin{eexample}\label{exampleCl30}
  {\it Exponent in \cl{3}{0}.}  Let's take   MV
  $\m{A}=-1+2 \e{1}-2 \e{12}-\e{123}-2 \e{13}+\e{2}+\e{23}+2 \e{3}$ the exponential of  which typifies non-diagonalizable MV.
  The minimal polynomial of  $\m{A}$ is
  $\mu_{\m{A}}(x)=\bigl(x^2+2 x+2\bigr)^2=x^4+4 x^3+8 x^2+8 x+4$ with $C_{(4)}= 4,C_{(3)}= 8,C_{(2)}=
   8,C_{(1)}= 4,C_{(0)}= 1$. It has two complex ($r\in \{1,2\}$) roots, namely, $\{x_1=\lambda_1=-1-\ii,x_2=\lambda_2= -1+\ii\}$.
   Each root has multiplicity two: $m_1=2$, and $m_2=2$. To avoid confusion, in the formulas below we have replaced
   $\lambda$ by $x$.  First, we will compute polynomials $p_1,q_1,p_2,q_2$ using a classical method described
   in Section~\ref{sec:classical}. In this particular case one finds two pairs of polynomials $\bigl\{
  \underbrace{p_1,q_1}_{m_1},  \underbrace{p_2,q_2}_{m_2} 
\bigr\}$ of multiplicity $m_1=m_2=2$.

Partial fraction decomposition of the first ($r=1$) root,
$\lambda_1=-1-\ii$, yields $p_1=\frac{1}{4} \ii (x+(1-\ii))^2
(x+(1+2 \ii))$ and $q_1=-\frac{1}{4} (x+(1-\ii))^2 (x+(1+\ii))$.
For the second ($r=2$) root $\lambda_2$ the polynomial is $p_2=-\frac{1}{4} \ii
(x+(1+\ii))^2 (x+(1-2 \ii))$ and $q_2=-\frac{1}{4} (x+(1-\ii))
(x+(1+\ii))^2$. Here we will only check that the polynomials indeed have
properties listed in Section~\ref{sec:classical}. For computational
details of $p_1,q_1,p_2,q_2$ the reader should refer to paper~\cite{Sobczyk1997}.  It is a straightforward
matter to check that $p_1+p_2=1$. The verification of multiplicative
properties requires division of polynomials modulo minimal polynomial. For example, to verify
that $p_1 q_1=q_1$ we have to divide the product $p_1 q_1$ by minimal polynomial
  $\mu_{\m{A}}(x)$, i.e. find $s(x)$ and $r(x)$ that allow to write the product in the form  $p_1(x) q_1(x)=
s(x)\mu(x) +r(x)$. Here $s(x)$ is the polynomial of lower degree, which after multiplication by minimal polynomial $\mu(x)$ and
summation with the reminder $r(x)$ yields the product $p_1(x)
q_1(x)$. Division by modulo minimal polynomial $\mu(x)$, denoted $\bigl( p_1(x) q_1(x)
  \mod \mu(x)\bigr)= r(x)$, means that we are
only interested in (keep) the reminder $r(x)$ and discard
  $s(x)\mu(x)$ part. In particular, to check the property $q_1^2=0$ we have to
  verify that $\bigl(q_1(x)^2 \mod \mu(x)\bigr) =0$, i.e.
  that $\mu(x)$ divides $q_1(x)^2$ (the reminder is zero). This is easy to do using computer algebra programs.
For example in {\it Mathematica}, this can be done with command {\bf
PolynomialReduce[}$p_1(x)\, q_1(x),\mu(x), x${\bf ]}. Then
  it is easy to check that $\bigl(p_1(x) q_1(x)\mod \mu(x)\bigr) =q_1(x)$
  and $\bigl(p_1(x) p_2(x)\mod \mu(x)\bigr) =0$. The calculations are similar for
polynomials $p_2(x), q_2(x)$ of the second root $\lambda_2$.

Once the generalized spectral basis was found, it is easy to check that the powers of MV satisfy the identities,
\begin{align*}
  \m{A}=&\bigl(\lambda_1+q_1(\m{A})\bigr) p_1(\m{A})+\bigl(\lambda_2+q_2(\m{A})\bigr) p_2(\m{A})\\
  \m{A}^2=&\bigl(\lambda_1+q_1(\m{A})\bigr)^2 p_1(\m{A})+\bigl(\lambda_2+q_2(\m{A})\bigr)^2 p_2(\m{A})\\
  \vdots&
\end{align*}
where  commutative variable $x$ was replaced by multivector  $\m{A}$.
The same replacement, which lies in the core of the Cayley-Hamilton theorem, was considered in Section~\ref{sec:charpoly}.

 Expressions for MV powers permit straightforward calculation of MV function by formula~\eqref{gNotation}, for example, to compute 
 the exponential $\exp(\m{A})$ we just need to replace the function $g$ by $\exp$. In the case of defective MV,
 the formula~\eqref{gNotation} requires computation of derivatives.
 Since in our case the roots have multiplicity $m_{1,2}=2$, only value of the first scalar derivative of the $\exp$
 function is required at roots $\lambda_{1,2}$. It is convenient at first to calculate all needed derivatives
 symbolically for arbitrary $\lambda$, and then substitute particular root $\lambda_i$.
 For $\exp(\lambda)$ the derivative is
  $\exp^\prime(\lambda_{i})=\frac{1}{1!} \frac{\dd\exp(\lambda+\tau)}{\dd \tau}\vert_{\tau=0,\lambda=\lambda_{i}}=\exp(\lambda_{i})$.
 Explicit formula in this case is
 \begin{align*}
   \exp(\m{A})=&\bigl(\exp(\lambda_1) p_1(\m{A})+\exp^\prime(\lambda_{1})q_1(\m{A})\bigr) +\bigl(\exp(\lambda_2) p_2(\m{A})+\exp^\prime(\lambda_{2})q_2(\m{A})\bigr)
\end{align*}
After substitution of $\m{A}=-1+2 \e{1}-2 \e{12}-\e{123}-2\e{13}+\e{2}+\e{23}+2 \e{3}$ one obtains
\begin{align*}
  \exp(\m{A})=&+\frac{1}{2} \left(1+\ee^{2 \ii}\right) \ee^{-1-\ii}+\frac{1}{2} \left((2+\ii)+(2-\ii) \ee^{2 \ii}\right) \ee^{-1-\ii} \e{1}+\frac{1}{2} \left((1+2 \ii)+(1-2 \ii) \ee^{2 \ii}\right) \ee^{-1-\ii} \e{2}+(1+\ii) \left(\ee^{2 \ii}-\ii\right) \ee^{-1-\ii} \e{3}
  \\&
-(1-\ii) \left(\ii+\ee^{2 \ii}\right) \ee^{-1-\ii} \e{12}-\frac{1}{2}
   \left((2-\ii)+(2+\ii) \ee^{2 \ii}\right) \ee^{-1-\ii} \e{13}+\frac{1}{2} \left((1-2 \ii)+(1+2 \ii) \ee^{2 \ii}\right)
   \ee^{-1-\ii} \e{23}\\&
+\frac{1}{2} \ii \left(-1+\ee^{2 \ii}\right) \ee^{-1-\ii} \e{123}.
\end{align*}
After summing complex conjugate roots pairwise, the answer can be rewritten in a real form,
  \begin{align*}
    \exp(\m{A})=\frac{1}{\ee}\Bigl(&
    \cos (1) + \e{1} (\sin (1)+2 \cos (1)) +\e{2} (2 \sin (1)+\cos (1)) +2 \e{3} (\cos (1)-\sin (1))\\&
    -2 \e{12} (\sin (1)+\cos (1))-\sin (1) \e{123}+\e{13} (\sin (1)-2 \cos (1))+\e{23} (\cos (1)-2 \sin (1))
    \Bigr)\,.
  \end{align*}

The essential part of the above computation (Section~\ref{sec:classical}) is the partial
fraction decomposition of $1/\mu(A)$, which yields the polynomials $p_i=Q_i^{0},q_i^1=Q_i^{1}, q_i^2=Q_i^{2},\cdots,q_i^{m_i-1}=Q_i^{m_i-1}$
and which usually is a nontrivial task. Even so, we shall demonstrate how easily the required
polynomials  can be obtained by recursion procedure~\eqref{recursionQ} from the main
Proposition~\ref{theoremMain}.

The first task is to calculate the largest polynomial power $Q_i^{m_i-1}(\m{A},\lambda)$ for each root (the
highest weight vector) that correspond to $q_i^{m_i-1}$ from classical method description in Section~\ref{sec:classical}.  Since the roots have
the multiplicity  $2$, the first line of the system~\eqref{recursionQ} yields
\begin{align*}
  &Q^{1}_{i=1,2}(x,\lambda)=\frac{S^{(0)}(x,\lambda)}{\frac{1}{2!}\mu^{\prime\prime}(\lambda)}=\frac{C_{(0)} x^3+x^2 (C_{(0)} \lambda +C_{(1)})+x \left(C_{(0)} \lambda ^2+C_{(1)} \lambda +C_{(2)}\right)+C_{(0)} \lambda ^3+C_{(1)} \lambda ^2+C_{(2)} \lambda +C_{(3)}}{4 \left(\lambda ^2+2 \lambda +2\right)+2 (2 \lambda +2)^2}
  \\&\hphantom{Q^{1}_{i=1,2}(x,\lambda)}=\frac{x^3+(\lambda +4) x^2+\left(\lambda ^2+4 \lambda +8\right) x+\lambda ^3+4 \lambda ^2+8 \lambda +8}{4 \left(\lambda ^2+2 \lambda +2\right)+2 (2 \lambda +2)^2},
  \\&Q^{1}_1(x,-1-\ii)=-\frac{1}{4} (x+(1-\ii))^2 (x+(1+\ii)),
\\&Q^{1}_2(x,-1+\ii)=-\frac{1}{4} (x+(1-\ii)) (x+(1+\ii))^2.
\end{align*}
  The polynomials $Q_i^{0}(x,\lambda)$, which represent $p_i$ in the classical method, can be found
  from the second line of the system~\eqref{recursionQ}.

\begin{align*}
  &Q^{0}_{i=1,2}(x,\lambda)=\frac{\frac{1}{1!}\frac{\dd S^{(0)}(x,\lambda)}{\dd \lambda}- Q^{1}_{i=1,2}(x,\lambda)\frac{1}{3!}\frac{\dd^3 \mu(\lambda)}{\dd \lambda^3}}{\frac{1}{2!}\frac{\dd^2 \mu(\lambda)}{\dd \lambda^2}} ={\textstyle\frac{(-2 \lambda -2) x^3+\left(\lambda ^2-4 \lambda -4\right) x^2+\left(4 \lambda ^3+14 \lambda ^2+8 \lambda \right)
   x+7 \lambda ^4+32 \lambda ^3+60 \lambda ^2+48 \lambda +16}{2 \left(3 \lambda ^2+6 \lambda +4\right)^2}},
  \\&Q^{0}_{1}(x,-1-\ii)= \frac{1}{4} \ii (x+(1-\ii))^2 (x+(1+2 \ii)),
\\&Q^{0}_{2}(x,-1+\ii)=-\frac{1}{4} \ii (x+(1+\ii))^2 (x+(1-2 \ii))\,.
\end{align*}
A bonus is that we have avoided polynomial division modulus minimal polynomial,
since the recursive computation starts from the "opposite end" as compared to  classical method.
Therefore, here proposed method is much simpler than that presented in detail in paper~\cite{Sobczyk1997},
which in turn may be advantageous over other known methods.
\end{eexample}

\section{Example: defective  MV  in \cl{4}{2}}\label{sec:exampleCl42}

To show how well the recursive formulas work, we will take a rather complicated MV which corresponds to a real matrix of
dimension $8\times 8$ and compute exponential function of the MV.
\begin{eexample}\label{exampleCl42}
  {\it Exponential in \cl{4}{2}.}  Let's take rather complicated and non-diagonalizable MV
  $\m{A}=\frac{1}{8} (2 \e{1}-\e{13}-\e{134}+2 \e{1345}-10 \e{13456}+4 \e{135}+2 \e{136}-4 \e{14}+\e{145}-2 \e{1456}+2 \e{146}-\e{15}+4 \e{16}-2 \e{34}-4 \e{345}+2
   \e{3456}-\e{346}-2 \e{35}-4 \e{356}+\e{36}+\e{456}+2 \e{5}+\e{56}+2 \e{6}+30)$
  the minimal polynomial of  which is
  $\mu_{\m{A}}(x)=(x-5)^4 (x-3)^3 (x-1)=x^8-30 x^7+386 x^6-2774 x^5+12132 x^4-32890 x^3+53550 x^2-47250 x+16875$ with $C_{(8)}= 16875$, $C_{(7)}= -47250$, $C_{(6)}= 53550$, $C_{(5)}= -32890$, $C_{(4)}= 12132$, $C_{(3)}= -2774$, $C_{(2)}= 386$, $C_{(1)}= -30$, and $C_{(0)}= 1$.

The MV has three real ($r\in \{1,2,3\}$) roots, namely, $\{x_1=\lambda_1=1,x_2=\lambda_2= 3, x_3=\lambda_3= 5 \}$. Corresponding multiplicities are  $m_1=1$, and $m_2=3$ and $m_3=4$. We need to compute polynomials $\bigl\{
  \underbrace{Q_1^0}_{m_1},\quad  \underbrace{Q_2^0,Q_2^1,Q_2^2}_{m_2},\quad \underbrace{Q_3^0,Q_3^1,Q_3^2,,Q_3^3}_{m_3}\quad
\bigr\}$.

First, we compute all parts that might enter into recursive formulas up to maximal multiplicity $m_i=4$.
The weighted derivatives of minimal polynomial are
  \begin{align}
    {\mu^{(1)}}(\lambda)=& 8 \lambda^7-210 \lambda^6+2316 \lambda^5-13870 \lambda^4+48528 \lambda^3-98670 \lambda^2+107100 \lambda -47250,\notag\\
     {\mu^{(2)}}(\lambda)=&28 \lambda ^6-630 \lambda ^5+5790 \lambda ^4-27740 \lambda ^3+72792 \lambda ^2-98670 \lambda +53550,\notag\\
      {\mu^{(3)}}(\lambda)=&56 \lambda ^5-1050 \lambda ^4+7720 \lambda ^3-27740 \lambda ^2+48528 \lambda -32890,\notag\\
       {\mu^{(4)}}(\lambda)=&70 \lambda ^4-1050 \lambda ^3+5790 \lambda ^2-13870 \lambda +12132, \\
        {\mu^{(5)}}(\lambda)=&56 \lambda ^3-630 \lambda ^2+2316 \lambda -2774, \notag\\
     {\mu^{(6)}}(\lambda)=&28 \lambda ^2-210 \lambda +386, \notag\\
      {\mu^{(7)}}(\lambda)=& 8 \lambda -30\notag\,.
\end{align}
  Polynomials $S^{(k)}(\m{A},\lambda)$ are\footnote{We computed all polynomials using commutative dummy variable, which in the last step was substituted by MV $\m{A}$ to keep formulas consistent with the form given in Proposition~\ref{theoremMain}. }

\begin{align}
  S^{(0)}(\m{A},\lambda)=&\m{A}^7-30 \m{A}^6+386 \m{A}^5-2774 \m{A}^4+12132 \m{A}^3+\left(\m{A}^2-30 \m{A}+386\right) \lambda ^5-32890 \m{A}^2+\left(\m{A}^3-30 \m{A}^2+386 \m{A}-2774\right) \lambda ^4\notag\\
  &+\left(\m{A}^4-30 \m{A}^3+386 \m{A}^2-2774
   \m{A}+12132\right) \lambda ^3+\left(\m{A}^5-30 \m{A}^4+386 \m{A}^3-2774 \m{A}^2+12132 \m{A}-32890\right) \lambda ^2\notag\\
  &+\left(\m{A}^6-30 \m{A}^5+386 \m{A}^4-2774 \m{A}^3+12132 \m{A}^2-32890 \m{A}+53550\right) \lambda
   +(\m{A}-30) \lambda ^6+53550 \m{A}+\lambda ^7-47250,\notag \\[4pt]
 S^{(1)}(\m{\m{A}},\lambda)= &\m{A}^6-30 \m{A}^5+386 \m{A}^4-2774 \m{A}^3+5 \left(\m{A}^2-30 \m{A}+386\right) \lambda ^4+12132 \m{A}^2+4 \left(\m{A}^3-30 \m{A}^2+386 \m{A}-2774\right) \lambda ^3\notag\\
  &+3 \left(\m{A}^4-30 \m{A}^3+386 \m{A}^2-2774
   \m{A}+12132\right) \lambda ^2+2 \left(\m{A}^5-30 \m{A}^4+386 \m{A}^3-2774 \m{A}^2+12132 \m{A}-32890\right) \lambda\notag\\
   &+6 (\m{A}-30) \lambda ^5-32890 \m{A}+7 \lambda ^6+53550,\notag\\[4pt]
  S^{(2)}(\m{\m{A}},\lambda)= &\m{A}^5-30 \m{A}^4+386 \m{A}^3+10 \left(\m{A}^2-30 \m{A}+386\right) \lambda ^3-2774 \m{A}^2+6 \left(\m{A}^3-30 \m{A}^2+386 \m{A}-2774\right) \lambda ^2\notag\\
  &+3 \left(\m{A}^4-30 \m{A}^3+386 \m{A}^2-2774 \m{A}+12132\right) \lambda
   +15 (\m{A}-30) \lambda ^4+12132 \m{A}+21 \lambda ^5-32890,\notag\\[4pt]
   S^{(3)}(\m{\m{A}},\lambda)= &\m{A}^4-30 \m{A}^3+10 \left(\m{A}^2-30 \m{A}+386\right) \lambda ^2+386 \m{A}^2+4 \left(\m{A}^3-30 \m{A}^2+386 \m{A}-2774\right) \lambda +20 (\m{A}-30) \lambda ^3\notag\\
   &-2774 \m{A}+35 \lambda ^4+12132\, .\notag
\end{align}

  Now for each root let's compute $Q^{k}_{i}(\m{A},\lambda)$ polynomials. Start from the root $\lambda=5$ the number of which is $i=3$ and the multiplicity is $m_3=4$ (the largest).
  One has to compute four polynomials,
\begin{align}
  {Q^{3}_3}(\m{A},5)=&\frac{1}{32} (\m{A}-5)^3 (\m{A}-3)^3 (\m{A}-1),\notag\\
 {Q^{2}_3}(\m{A},5)=&-\frac{1}{128} (\m{A}-5)^2 (\m{A}-3)^3 (\m{A}-1) (7 \m{A}-39),\notag\\
 {Q^{1}_3}(\m{A},5)=&\frac{1}{512} (\m{A}-5) (\m{A}-3)^3 (\m{A}-1) \left(31 \m{A}^2-338 \m{A}+931\right), \notag\\
 {Q^{0}_3}(\m{A},5)=&-\frac{1}{2048}(\m{A}-3)^3 (\m{A}-1) \left(111 \m{A}^3-1789 \m{A}^2+9677 \m{A}-17599\right)\notag\,.
\end{align}

The same polynomials for root $\lambda=3$, which number is $i=2$ and the multiplicity  $m_2=3$ are
\begin{align}
 {Q^{2}_2}(\m{A},3)=&\frac{1}{32} (\m{A}-5)^4 (\m{A}-3)^2 (\m{A}-1),\notag\\
 {Q^{1}_2}(\m{A},3)=&\frac{1}{64} (\m{A}-5)^4 (\m{A}-3) (\m{A}-1) (3 \m{A}-7), \notag\\
 {Q^{0}_2}(\m{A},3)=&\frac{1}{128} (\m{A}-5)^4 (\m{A}-1) \left(7 \m{A}^2-36 \m{A}+49\right)\notag\,.
\end{align}
For the first  $i=1$  root $\lambda=1$ with multiplicity $m_1=1$ we need only
  \begin{align}{Q^{0}_1}(\m{A},1)=&-\frac{(\m{A}-5)^4 (\m{A}-3)^3}{2048}\,.
   \end{align}

This ends up computation of generalized spectral basis. Before going over to computation of exponential,
let's check the identity $\m{A}=1\ {Q^{0}_1}(\m{A},1) + \bigl(3 + {Q^{1}_2}(\m{A},3)\bigr)
  {Q^{0}_2}(\m{A},3) + \bigl(5 +  {Q^{1}_3}(\m{A},5)\bigr){Q^{0}_3}(\m{A},5)
$.  Substitution of all $Q^{k}_j$ into RHS yields
  \begin{align}&\frac{1}{1048576}  \bigl(-753 \m{A}^{14}+37946 \m{A}^{13}-874979 \m{A}^{12}+12226436 \m{A}^{11}-115570953 \m{A}^{10}+781115526 \m{A}^9-3889496339 \m{A}^8\notag\\
    &+14481221176 \m{A}^7-40461630635 \m{A}^6+84287184806 \m{A}^5-128533325601
    \m{A}^4+138586998020 \m{A}^3-99271453275 \m{A}^2\notag\\
    &+41924292826 \m{A}-7799675625\bigr)\,,
   \end{align}
Substituting initial MV $\m{A}$ into latter expression and expanding all geometric powers
one indeed obtains $\m{A}$, which is the LHS of the identity.

Similar check of identity $\m{A}^2=1^2\ {Q^{0}_1}(\m{A},1) + \bigl(3 + {Q^{1}_2}(\m{A},3)\bigr)^2
  {Q^{0}_2}(\m{A},3) + \bigl(5 +  {Q^{1}_3}(\m{A},5)\bigr)^2{Q^{0}_3}(\m{A},5)
$ yields true answer too. This indicates  that generalized spectral basis was computed correctly and
we can proceed with computation of MV function using formula~\eqref{gTaylorFin}.
We only need to compute some derivatives of exponential function. Then, resorting to already described procedure
we  get  final answer for MV exponential function,
\begin{align}
  \exp(\m{A})=&-\frac{1}{48} \ee \Bigl(\left(6-6 \ee^2\right) \e{1}+6 \left(\ee^2+\ee^4\right) \e{13}+6 \left(1+\ee^2-2 \ee^4\right) \e{1345}+6 \left(-1-3 \ee^2+4 \ee^4\right) \e{13456}+\ee^2 \left(7
   \ee^2-6\right) \e{1346} \notag\\
   &-6 \left(-1+\ee^2+2 \ee^4\right) \e{135}+6 \left(1+\ee^2\right) \e{1356}-5 \ee^4 \e{136}+\left(18 \ee^2-6\right) \e{14}+\ee^2 \left(6+7 \ee^2\right)
   \e{1456}+\left(6-6 \ee^2\right) \e{146}\notag\\
   &-6 \ee^2 \left(\ee^2-1\right) \e{15}+5 \ee^4 \e{156}+\left(6-18 \ee^2\right) \e{16}+\ee^2 \left(7 \ee^2-6\right) \e{3}+5 \ee^4 \e{34}+\left(18
   \ee^2-6\right) \e{345}+\left(6-6 \ee^2\right) \e{3456}\notag\\
   &+6 \left(\ee^2+\ee^4\right) \e{346}+6 \left(\ee^2-1\right) \e{35}+\left(18 \ee^2-6\right) \e{356}+6 \left(1+\ee^2\right)
   \e{4}+5 \ee^4 \e{45}+6 \ee^2 \left(\ee^2-1\right) \e{456}\notag\\
  &+6 \left(-1+\ee^2-2 \ee^4\right) \e{46}-\ee^2 \left(6+7 \ee^2\right) \e{5}-6 \left(1+\ee^2+2 \ee^4\right) \e{6}-6 \left(1+3
   \ee^2+4 \ee^4\right)\Bigr)\notag\,.
\end{align}
The answer was checked that it satisfies a defining property of MV exponential, Eq.~\eqref{expproperty}.
\end{eexample}

\section{Example: defective  MV  in \cl{4}{2} with high degree irreducible polynomial}\label{sec:exampleImplicit}
The last example demonstrates that for some inputs even sophisticated computer algebra systems can have problems computing functions of matrices, where our method handles the cases efficiently and flawlessly. Unfortunately, the computed answers are too large to be presented explicitly, therefore we will compare results by evaluating timing and complexity of answers by counting number of leafs of returned symbolic expressions.

\begin{eexample}
 To illustrate the point, in $\cl{4}{2}$ algebra we take a
 non-diagonalizable MV,
 \begin{align}
   \m{A}=&
-1-\e{3}+\e{6}-\e{12}-\e{13}+\e{15}-\e{24}-\e{25}+\e{26}-\e{34}-\e{35}+\e{36}-\e{45}+\e{56}+\e{123}+\e{124}+\e{126}+\e{134}\notag\\
   &+\e{135}+\e{136}+\e{146}+\e{234}-\e{235}-\e{236}-\e{245}-\e{246}-\e{256}+\e{456}
-\e{1236}+\e{1245}-\e{1246}+\e{1256}-\e{1345}\\
   &-\e{1346}-\e{1356}+\e{1456}-\e{2346}-\e{2356}+\e{2456}+\e{3456}+\e{12345}-\e{12346}+\e{12356}\,.\notag
  \end{align}
and compute $\exp(\m{A})$.

The MV has the following $8\times 8$ real matrix representation
\begin{align}
 \left(\begin{matrix}
  0&0&0&-2&-2&0&2&-1\\
  0&-2&4&2&-2&2&5&2\\
  4&0&0&2&4&-3&-2&6\\
  -2&-2&2&-2&-1&-4&0&2\\
  0&-2&2&-1&0&2&-2&-2\\
  0&-4&1&2&-2&2&-4&0\\
  0&1&0&0&2&0&-4&0\\
  -1&0&-2&4&-2&4&-4&-2
\end{matrix}\right)
\end{align}
\noindent
in $\cl{4}{2}$ algebra basis that is expressed by matrices 
\begin{align}
&\hat{\e{}}_1= \left(
\begin{smallmatrix}
 0 & 1 & 0 & 0 & 0 & 0 & 0 & 0 \\
 1 & 0 & 0 & 0 & 0 & 0 & 0 & 0 \\
 0 & 0 & 0 & -1 & 0 & 0 & 0 & 0 \\
 0 & 0 & -1 & 0 & 0 & 0 & 0 & 0 \\
 0 & 0 & 0 & 0 & 0 & 1 & 0 & 0 \\
 0 & 0 & 0 & 0 & 1 & 0 & 0 & 0 \\
 0 & 0 & 0 & 0 & 0 & 0 & 0 & -1 \\
 0 & 0 & 0 & 0 & 0 & 0 & -1 & 0 \\
\end{smallmatrix}
\right), \hat{\e{}}_2= \left(
\begin{smallmatrix}
 0 & 0 & 0 & 0 & 0 & 0 & 0 & -1 \\
 0 & 0 & 0 & 0 & 0 & 0 & -1 & 0 \\
 0 & 0 & 0 & 0 & 0 & -1 & 0 & 0 \\
 0 & 0 & 0 & 0 & -1 & 0 & 0 & 0 \\
 0 & 0 & 0 & -1 & 0 & 0 & 0 & 0 \\
 0 & 0 & -1 & 0 & 0 & 0 & 0 & 0 \\
 0 & -1 & 0 & 0 & 0 & 0 & 0 & 0 \\
 -1 & 0 & 0 & 0 & 0 & 0 & 0 & 0 \\
\end{smallmatrix}
\right), \hat{\e{}}_3= \left(
\begin{smallmatrix}
 0 & 0 & 0 & -1 & 0 & 0 & 0 & 0 \\
 0 & 0 & -1 & 0 & 0 & 0 & 0 & 0 \\
 0 & -1 & 0 & 0 & 0 & 0 & 0 & 0 \\
 -1 & 0 & 0 & 0 & 0 & 0 & 0 & 0 \\
 0 & 0 & 0 & 0 & 0 & 0 & 0 & 1 \\
 0 & 0 & 0 & 0 & 0 & 0 & 1 & 0 \\
 0 & 0 & 0 & 0 & 0 & 1 & 0 & 0 \\
 0 & 0 & 0 & 0 & 1 & 0 & 0 & 0 \\
\end{smallmatrix}
\right),\notag\\
&\hat{\e{}}_4= \left(
\begin{smallmatrix}
 1 & 0 & 0 & 0 & 0 & 0 & 0 & 0 \\
 0 & -1 & 0 & 0 & 0 & 0 & 0 & 0 \\
 0 & 0 & 1 & 0 & 0 & 0 & 0 & 0 \\
 0 & 0 & 0 & -1 & 0 & 0 & 0 & 0 \\
 0 & 0 & 0 & 0 & 1 & 0 & 0 & 0 \\
 0 & 0 & 0 & 0 & 0 & -1 & 0 & 0 \\
 0 & 0 & 0 & 0 & 0 & 0 & 1 & 0 \\
 0 & 0 & 0 & 0 & 0 & 0 & 0 & -1 \\
\end{smallmatrix}
\right), \hat{\e{}}_5= \left(
\begin{smallmatrix}
 0 & 0 & 0 & 1 & 0 & 0 & 0 & 0 \\
 0 & 0 & 1 & 0 & 0 & 0 & 0 & 0 \\
 0 & -1 & 0 & 0 & 0 & 0 & 0 & 0 \\
 -1 & 0 & 0 & 0 & 0 & 0 & 0 & 0 \\
 0 & 0 & 0 & 0 & 0 & 0 & 0 & 1 \\
 0 & 0 & 0 & 0 & 0 & 0 & 1 & 0 \\
 0 & 0 & 0 & 0 & 0 & -1 & 0 & 0 \\
 0 & 0 & 0 & 0 & -1 & 0 & 0 & 0 \\
\end{smallmatrix}
\right),\qquad\hat{\e{}}_6= \left(
\begin{smallmatrix}
 0 & 1 & 0 & 0 & 0 & 0 & 0 & 0 \\
 -1 & 0 & 0 & 0 & 0 & 0 & 0 & 0 \\
 0 & 0 & 0 & 1 & 0 & 0 & 0 & 0 \\
 0 & 0 & -1 & 0 & 0 & 0 & 0 & 0 \\
 0 & 0 & 0 & 0 & 0 & 1 & 0 & 0 \\
 0 & 0 & 0 & 0 & -1 & 0 & 0 & 0 \\
 0 & 0 & 0 & 0 & 0 & 0 & 0 & 1 \\
 0 & 0 & 0 & 0 & 0 & 0 & -1 & 0 \\
\end{smallmatrix}
\right).
\end{align}

The respective MV/matrix has minimal polynomial $(\lambda -1)^2 \left(\lambda ^6+10 \lambda ^5+39 \lambda
^4+124 \lambda ^3+543 \lambda ^2-198 \lambda -4743\right)$. 
  While {\it Mathematica} version 13.0 in our previous experiments~\cite{AcusMMAS2023} 
  has crashed on this input after 48 hours of computation after it has exhausted all 96GB of server RAM, the version 14.1 successfully completes the same task on laptop with 16GB RAM in about 64 seconds. 
  This is to be compared to 1.5 seconds for the same calculations using our method for MV or just to 0.5 second when using matrices (the difference is due to highly optimized multiplication of matrices in {\it Mathematica}) on the same hardware. The leaf count (measure of complexity) of the result also differs by orders of magnitude. Leaf count of our complex valued result (answer in real form is twice smaller) is less than $10^6$ (when converted to matrix form for comparison), while  leaf count of 
\textit{Mathematica} function {\bf MatrixFunction[~]} output\footnote{Specialized function {\bf MatrixExp[]} is slightly slower than {\bf MatrixFunction[~]} and returns the same answer.} is
almost $2.5\times 10^9$. Even numerical comparison of both matrices up to 100 number digit precision took over 780 seconds
(numerical evaluation time of our output to this precision is negligible, 1/10 second). 
\end{eexample}

\section{Conclusions and perspective}
\label{sec:discussion}

The paper provides a recursive method to compute generalized spectral basis of multivectors (MVs) in Clifford geometric
 algebra.  The method can be easily extended to the square defective (non-diagonalizable) matrices as well.
 The generalized spectral basis opens fast and easy way to compute functions of the MVs, matrices and linear operators.
 The main result of this paper is presented in Proposition~\ref{theoremMain} (formulas
 \eqref{funcCoordFree}, \eqref{recursionQ} and \eqref{formulasS}). Using  defective MVs a number of examples with real MVs
 are presented that demonstrate  how the method works  in practice. Also comparison with the classical method is provided.
 As far as we know the proposed recursive method provides \textit{exact}, simplest,  and fastest  tool
 to compute functions of MV and matrix as well, since  computation of the generalized
 spectral  basis does not require polynomial reduction. At present, approximate numerical
 methods are mainly used to evaluate functions of matrix~\cite{Higham2008}.  However, if exact results are need,
 even the sophisticated computer algebra systems  have problems in computing exact MV/matrix function of quite simple form,
 when minimal polynomial includes  a high degree (irreducible) polynomial as was demonstrated in Section~\ref{sec:exampleImplicit}.

Though at the moment we are unable to provide a strict proof of Proposition~\ref{theoremMain} for the  most general case, we have no doubts that
 our presented recursive method provides better and more preferable way to compute generalized spectral basis and, in its turn,
 gives the go-ahead to do calculations with  functions of MV/matrix argument. In fact, the definition of minimal polynomial of MV and  existence of new
recursive sequence itself provides an exciting research and application perspective. For example, recently there were attempts~\cite{ShirokovSVG2024,ShirokovRank2024} to define MV rank by singular value decomposition of MV without any reference to matrix representations. 
This work suggests that the rank of MV can be defined simply as a degree of minimal polynomial,  $\textrm{rank}(A)=\mathrm{deg}(\mu(\m{A}))=\sum_{i=0}^r m_i$ (or may be as a difference between highest and smallest exponents of monomials) and computed by Algorithm~\ref{algMinimalPoly}, without any reference to matrix representation as well. 

To summarise, we have employed a number of examples using exact
numerical MVs/matrices to demonstrate how the method works. The recursive generalized spectral basis computation procedure can be applied to matrices of general linear
group/algebra, i.e. to matrices (with symbolic entries as well) without any restrictions or conditions on its elements, and there are plenty problems in physics that require to find  functions of matrices, for which close form solutions would be interesting to know. We hope that the  generalized spectral basis
will help to solve some of the mentions problems.

\subsection*{Conflict of interest}

The authors declare no potential conflict of interests.

\section*{Supporting information}

There are no funders to report for this submission.

\bibliography{expNDNondiag}%

\begin{thebibliography}{10}

\bibitem{Gurlebeck1997}
G{\"u}rlebeck K., Spr{\"o}ssig W.. {\it Quaternionic and {C}lifford Calculus
  for Physicists and Engineers}.
\newblock Chichester, England: John Wiley and Sons; 1998.
\newblock ISBN: 978-0-471-96200-7.

\bibitem{Lounesto1997}
Lounesto P.. {\it Clifford Algebra and Spinors}.
\newblock Cambridge: Cambridge University Press; 1997.
\newblock ISBN-13: 978-0521599160.

\bibitem{Marchuk2020}
Marchuk N.~G., Shirokov D.S.. {\it Theory of {C}lifford Algebras and Spinors}.
\newblock Moscow: Krasand; 2020.
\newblock ISBN: 978-5-396-01014-7, in Russian.

\bibitem{Higham2008}
Higham N.~J.. {\it Functions of Matrices (Theory and Computation)}.
\newblock Philadelphia: SIAM; 2008.
\newblock ISBN: 978-0-898716-46-7.

\bibitem{AcusMMAS2023}
Acus A., Dargys A.. The characteristic polynomial in calculation of exponential
  and elementary functions in {C}lifford algebras.  {\it Math. Meth. Appl.
  Sci.. }2023;:1-15.
\newblock doi.org/10.1002/mma.9524.

\bibitem{AcusDargys2024log}
Acus A., Dargys A.. Logarithm of multivector in real 3D Clifford algebras.
  {\it Nonlinear Analysis: Modelling and Control. }2024;29(1):13-31.

\bibitem{AcusDargys2024}
Acus A., Dargys A.. Coordinate-free exponentials of general multivector in
  Cl(p,q) algebras for p+q=3.  {\it Mathematical Methods in the Applied
  Sciences. }2024;47(3):1362-1374.

\bibitem{Sobczyk1997}
Sobczyk G.. The Generalized Spectral Decomposition of a Linear Operator.  {\it
  The College Mathematics Journal. }1997;28(1):27--38.

\bibitem{Costa2004OnTE}
Ramakrishna Viswanath. On the exponentials of some structured matrices.  {\it
  Journal of Physics A. }2004;37:11613-11627.

\bibitem{Zhou2005}
Ramakrishna Viswanath, Zhou Hong. On the Exponential of Matrices in su(4).
  {\it Journal of Physics A. }2005;39:3021-3034.
\newblock
  doi:10.1088/0305-4470/39/12/011,https://arxiv.org/abs/math-ph/0508018v1.

\bibitem{Fujii2007ExponentiationOC}
Fujii Kazuyuki. Exponentiation of certain Matrices related to the Four Level
  System by use of the Magic Matrix.  {\it arXiv:math-ph/0508018v1. }2007;.

\bibitem{Fujii2012}
Fujii Kazuyuki, Oike Hiroshi. How to Calculate the Exponential of Matrices.
  {\it Far East Journal of Mathematical Education. }2012;9(1):39-55.
\newblock P-ISSN: 0973-5631; arXiv:quant-ph/0604115v1.

\bibitem{Herzig2014}
Herzig Emily, Ramakrishna Viswanath, Dabkowski Mieczyslaw~K.. {Note on
  Reversion, Rotation and Exponentiation in Dimensions Five and Six}.  {\it
  Journal of Geometry and Symmetry in Physics. }2014;35(none):61 -- 101.

\bibitem{Zela2014}
Zela F.~De. Closed-Form Expressions for the Matrix Exponential.  {\it Symmetry.
  }2014;6:329-344.
\newblock ISSN 2073-8994. doi:10.3390/sym6020329,
  www.mdpi.com/journal/symmetry.

\bibitem{Householder1975}
Householder Alston~S.. {\it The Theory of Matrices in Numerical Analysis}.
\newblock New York, N. Y. 10014: Dover Publications, Inc; 1975.

\bibitem{Hou1998}
{Shui-Hung~Hou} . Classroom note: A Simple Proof of the {L}eVerrier-{F}addeev
  Characteristic Polynomial Algorithm.  {\it SIAM Review.. }1998;40(3):706-709.

\bibitem{Shirokov2021}
Shirokov D.~S.. On Computing the Determinant, other Characteristic Polynomial
  Coefficients, and Inverses in {C}lifford Algebras of Arbitrary Dimension.
  {\it Comput. Appl. Math.. }2021;40(173):1-29.
\newblock doi.org/10.1007/s40314-021-01536-0.

\bibitem{Garibaldi2004}
Garibaldi S.. The characteristic polynomial and determinant are not \textit{ad
  hoc} constructions.  {\it The American Mathematical Monthly.
  }2004;111(9):761-778.

\bibitem{Mathworld2022}
Rowland Todd, Weisstein Eric~W.. {\it Matrix Minimal Polynomial. } From
  MathWorld--A Wolfram Web Resource.
  \url{https://mathworld.wolfram.com/MatrixMinimalPolynomial.html}; 2025.

\bibitem{AcusDargys2025}
Acus A., Dargys A.. {\it {G}eometric {A}lgebra \textit{{M}athematica} package.
  } Download from \url{https://github.com/ArturasAcus/GeometricAlgebra}; 2025.

\bibitem{Georgi2000}
Georgi H.. {\it Lie Algebras In Particle Physics: from Isospin To Unified
  Theories}.
\newblock Boca Raton: CRC Press; 2000.

\bibitem{ShirokovSVG2024}
Shirokov D.. On SVD and Polar Decomposition in Real and Complexified Clifford
  Algebras.  {\it Adv. Appl. Clifford Algebras. }2024;34(23).

\bibitem{ShirokovRank2024}
Shirokov D.. On Rank of Multivectors in Geometric Algebras.  {\it
  arXiv:2412.02681v1. }2024;.

\end{thebibliography}

%
%

\end{document}